\documentclass[twocolumn,english,prl,showpacs,amsmath,amssymb]{revtex4}
\usepackage[T1]{fontenc}
\usepackage[latin9]{inputenc}
\usepackage{color}
\usepackage{graphicx}
\usepackage{babel}

\begin{document}
\title{Hydrodynamic Limit for the Spin Dynamics of the Heisenberg Chain}
\author{Simon Grossjohann and Wolfram Brenig}
\email{s-n.grossjohann@tu-bs.de}
\affiliation{
Institut f\"ur Theoretische Physik,
Technische Universit\"at Braunschweig,
38106 Braunschweig, Germany}

\begin{abstract}
We show that Quantum-Monte-Carlo calculations of the
dynamic structure factor of the isotropic spin-1/2 antiferromagnetic
chain at intermediate temperatures corroborate a picture of diffusive spin dynamics at finite frequencies in
the low-energy long wave-length limit and are in good agreement with recent
predictions for this by J. Sirker, R. G. Pereira, and I. Affleck [arXiv:0906.1978v1]. 
\end{abstract}

\maketitle
The one-dimensional (1D) Heisenberg $XXZ$ antiferromagnet
\begin{equation}
H=J\sum_{l}\left[\Delta S_{l}^{z}S_{l+1}^{z}+
\frac{1}{2}(S_{l}^{+}S_{l+1}^{-}+
S_{l}^{-}S_{l+1}^{+})\right]\,\,,
\label{1}
\end{equation}
where $J>0$ is the exchange coupling, $S_{l}^{z,\pm}$ are spin-$1/2$
operators on site $l$, and $\Delta$ is the exchange anisotropy is
one of the best studied strongly correlated many-body system. Its
magnetic transport properties however, remain an open issue \cite{fhm07}.
Spin transport in the Heisenberg chain is directly related to carrier
transport in 1D correlated spinless fermion systems, via the Jordan-Wigner
transformation, and therefore is of great interest in a broader context.
Linear response theory \cite{fhm03} shows the zero momentum, frequency
dependent spin conductivity
\begin{equation}
\sigma'(\omega)=D\delta(\omega)+\sigma'_{reg}(\omega)
\label{2}
\end{equation}
to consist of the Drude weight
\begin{equation}
D=\frac{\beta}{N}\sum_{\stackrel{m,n}{E_{m}\neq E_{n}}}e^{-\beta E_{m}}
|\langle m|j|n\rangle|^{2}
\label{3}
\end{equation}
and a regular spectrum
\begin{align}
\sigma'_{reg}(\omega) & =\frac{1-e^{-\beta\omega}}{\omega}
\frac{1}{N}\sum_{\stackrel{m,n}{E_{m}\neq E_{n}}}
\left[e^{-\beta E_{m}}\,\times\right.
\nonumber \\
& \,\,\,\,\,\left.|\langle m|j|n\rangle|^{2}
\delta(\omega-E_{n}+E_{m})\right]
\label{4}
\end{align}
where $j=j_{q=0}$ is the z-component of the spin current with
$j_{q}=(i\Delta J/2)\sum_{l}\exp(-iql)(S_{l}^{-}S_{l+1}^{+}-
S_{l}^{+}S_{l+1}^{-})$ and $m$, $n$ are the eigenstates with
energies $E_{m,n}$.

The Drude weight has been under intense scrutiny for more than two
decades. However, no generally accepted picture has emerged.
A nonzero Drude weight would imply dissipationless transport in a
correlated system \cite{zotos97}, despite the fact that $[j,H]\neq0$ for
the $XXZ$ model. Here we give a brief summary regarding the status of
this issue and refer to \cite{fhm07} and refs. therein for a more extensive summary. At $T=0$ and in the massless regime $|\Delta|<1$ of the $XXZ$ chain, the zero temperature Drude weight is known to be
finite \cite{shastry90}. At $T\neq0$, Bethe-Ansatz (BA) calculations
arrive at contradictory results regarding the temperature dependence
of $D(T)$ \cite{zotos99,benz05,lan07}. The same holds for the question
whether $D(T>0)$ is finite or not at the $SU(2)$ symmetric point
$\Delta=1$ \cite{zotos99,benz05}. Recent numerical studies using
QMC \cite{alvarez02,Heidarian07}, exact diagonalization (ED) at zero
\cite{fhm03,narozhny1998,jung2007,mukerjee08}, as well as finite magnetic
fields \cite{fhm05}, and master equations \cite{michel08,prosen09}
are consistent with $D\neq0$ for $|\Delta|\leq$1 and $T\ge0$,
supporting a ballistic contribution to the conductivity at
finite temperatures. Recent time-dependent density-matrix renormalization
group (tDMRG) studies have given evidence for ballistic spin dynamics
for $|\Delta|\leq$1 in the out-of-equilibrium case \cite{langer09}.

The regular finite-frequency contribution $\sigma'_{reg}(\omega)$ has
been considered by ED studies \cite{zotos96,naef98}, which
however leave many open issues. Very recently, spin diffusion has been
conjectured to govern the low-frequency spectrum of the regular conductivity
\cite{Sirker09}, based on real-time transfer matrix renormalization
group (tTMRG) and a perturbative analysis using bosonization. The
latter provides for an approximate expression for the Fourier transform
of the retarded spin-susceptibility $\chi_{ret}(q,t)= i\Theta(t)\langle[S_{q}^{z}(t),S_{-q}^{z}]\rangle$, which reads
\begin{equation}
\chi_{ret}\left(q,\omega\right)=
-\frac{Kvq^{2}}{2\pi}\frac{1}{\omega^{2}-v^{2}q^{2}-\Pi_{ret}
\left(q,\omega\right)}\,\,,
\label{5}
\end{equation}
with
\begin{equation}
\Pi_{ret}\left(q,\omega\right)\approx
-2i\gamma_{B}\omega-b\omega^{2}+cv^{2}q^{2}\,\,,
\label{6}
\end{equation}
where at $\Delta=1$, $K=1$, $v=\pi/2$, $2\gamma_{B}=\pi g^{2}T$,
$b=g^{2}/4-g^{3}(3-8\pi^{2}/3)/32+\sqrt{3}T^{2}/ \pi$, and 
$c=g^{2}/4-3g^{3}/32-\sqrt{3}T^{2}/\pi$ have been obtained
by perturbative expansions (PE) at $T\ll J$ \cite{Sirker09} 
in powers of the running coupling constant $1/g+\ln(g)/2=
\ln\left(\sqrt{\pi/2}\exp(G+1/4)/T\right)$ and $G\approx0.577216\ldots$
is Euler's constant \cite{Lukyanov98}.

Some remarks are in order. First, for $\omega\ll\gamma$, eqn. (\ref{5})
displays a diffusion pole with a diffusion constant $\Gamma=(1+c)v^{2}/
(\pi g^{2}T)$. I.e. within this approximation the spin dynamics of the
Heisenberg chain would allow for a plain hydrodynamic limit. Second, eqns.
(\ref{5}) and (\ref{6}) do not incorporate the finite width of the spectral
function $\chi''\left(q,\omega\right)=Im[\chi_{ret}(q,\omega)]/\pi$
at $T=0$, which is dominantly set by the two-spinon continuum. However,
at $q/\pi\ll1$ the latter width is of order $\pi Jq^{3}/16$, which
for those wave vectors and temperatures which we will be interested
in is negligible against $\gamma_{B}$. Third, for any finite
momentum $q\neq0$, the isothermal susceptibility $\chi_{q}=\int_{-\infty}^{\infty}d\omega\chi''(q,\omega)/\omega$
obtained from eqn. (\ref{5}) is identical to the isolated susceptibility
$\chi_{ret}(q,0)=\int_{-\infty}^{\infty}d\omega
\chi''(q,\omega)/(\omega-i0^{+})$,
since $\chi''(q\neq0,\omega\rightarrow0)\propto\omega$. Therefore
$\chi_{q}=K/(2\pi v(1+c))$. Furthermore, the isothermal susceptibility
of the Heisenberg model is a continuous function of $q$. Its limiting
value $\lim_{q\rightarrow0}\chi_{q}=\chi_{0}$ at zero momentum is known from
thermodynamic Bethe Ansatz (TBA) \cite{Lukyanov98,Johnston00}. Therefore
\begin{equation}
\frac{K/(2\pi)}{v(1+c)}=\chi_{0}\approx\frac{1}{\pi^{2}}
(1-\frac{g^{2}}{4}+\frac{3g^{3}}{32}+
\frac{\sqrt{3}T^{2}}{\pi})=\chi_{PE}\,\,\,,
\label{8}
\end{equation}
should be satisfied, where $\chi_{PE}$ is a known PE of the TBA result
\cite{Lukyanov98,Johnston00}. This implies, that apart from eqns.
(A2) and (B3) of ref. \cite{Sirker09}, also the ratio $K/v$ requires
to be renormalized off from $2/\pi$ \cite{Pcommun}.

The spectral function $\chi''(q,\omega)$ is related to
$\sigma'_{reg}(\omega)$ by means of the lattice version of
the continuity equation $\partial_{t}S_{q}^{z}=(1-\exp(-iq))\, j_{q}$
through
\begin{equation}
\sigma'_{reg}(\omega)=\lim_{q\rightarrow0}\frac{\omega}{q^{2}}
\chi''(q,\omega)\,\,.
\label{7}
\end{equation}
Therefore, the spectrum of the regular part of the optical conductivity
can be deduced from eqns. (\ref{5}) and (\ref{6}).

The main goal of this work is to analyze, to which extent eqns. (\ref{5})
and (\ref{6}) are consistent with QMC calculations. The significance of such comparison is with the regular part of the spin conductivity. It will {\em not} clarify the size
of the Drude weight, as any discrepancy arising may be due to partial
spectral weight transfer into a Drude weight. Furthermore, we focus on
the isotropic point $\Delta=1$, which may be different from
the anisotropic case.
To begin, we note, that eqns. (\ref{5}) and
(\ref{6}) approximate the on-shell part of the spectrum for $|\omega\pm vq|
\ll T$. Yet, similar to the comparison with tTMRG in eqns. (C2) and (C3) of
ref. \cite{Sirker09}, we will assume them to be valid for all $\omega$.
Furthermore, $\chi_{q}$ is known to monotonously increase for the
Heisenberg model as $q\rightarrow\pi/2$. However, $\chi_{q}=K/(2\pi v(1+c))$
from bosonization is momentum independent. Therefore, a momentum dependence
$K\rightarrow K_{q}$, $v\rightarrow v_{q}$ - albeit weak at $q\ll1$
- is to be allowed for, when matching up eqns. (\ref{5}) and (\ref{6})
with QMC.

\begin{figure}
\begin{centering}
\includegraphics[scale=0.56]{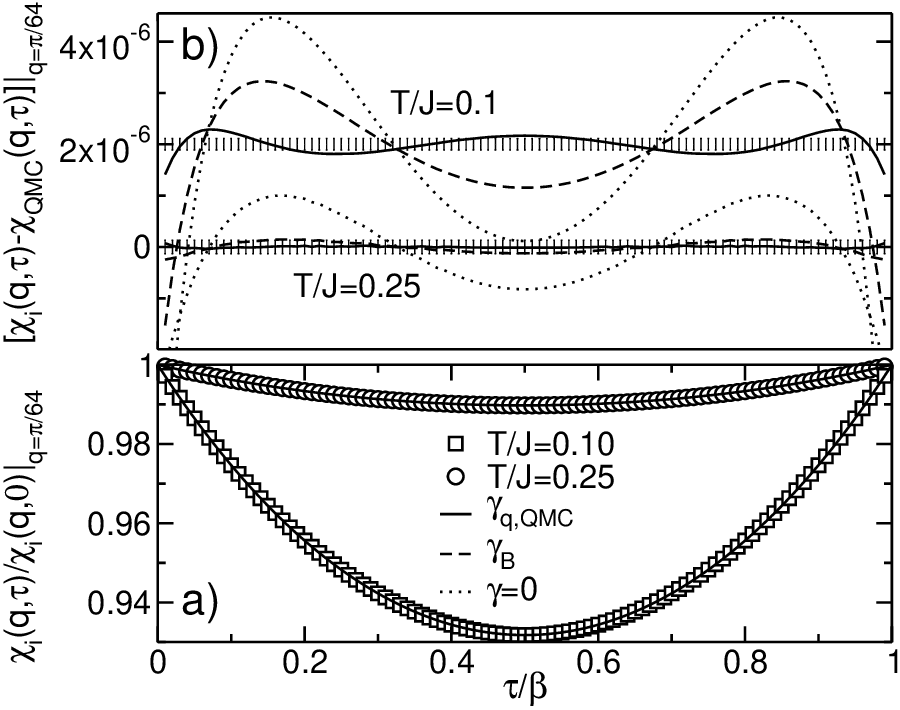}
\end{centering}

\caption{\label{fig:1}Imaginary-time susceptibility $\chi_{QMC}(q,\tau)$
at $q=\pi/64$ on 128 sites, for two temperatures $T$, fitted to
$\chi(q,\tau)$ from eqn. (\ref{10}) (lines) in three ways, namely:
$\gamma_{q,QMC}$ optimized (solid), $\gamma_{B}$ taken from
ref. \cite{Sirker09} (dashed), and $\gamma$ forced to zero (dotted).
The index 'i' on the y-axis refers to $\chi(q,\tau)$
from eqn. (\ref{10}) for the lines in panels a) and b) as well as to
QMC for the symbols in panel a).  Panel a) Global
behavior of $\chi_{QMC}(q,\tau)/\chi_{QMC}(q,0)$ for $T/J=0.1$ (QMC,
squares) and $0.25$ (QMC, circles). In this panel the three fits (lines) are
indistinguishable on the scale of the plot. Panel b) Error $2\sigma$
of $\chi_{QMC}(q,\tau)$ for each $\tau$ evaluated (error bars) and
difference $\chi(q,\tau)-\chi_{QMC}(q,\tau)$ between QMC and the
three fits (lines). $2\sigma$ for the QMC data is $O(10^{-7})$.
Plots corresponding to $T/J=0.1$ have been shifted by $2\times10^{-6}$.}
\end{figure}

We perform the comparison to QMC by transforming
$\chi_{ret}(q,\omega)$ onto the imaginary time axis
\begin{align}
\chi\left(q,\tau\right) & = & 2\sum_{n=0}^{\infty}
\cos(\omega_{n}\tau)\chi(q,\omega_{n})-\chi(q,0) \nonumber \\
\chi(q,\omega_{n}) & = & \frac{K_{q}v_{q}q^{2}/(2\pi)}{(1+b)
\omega_{n}^{2}+(1+c)v_{q}^{2}q^{2}+2\gamma_{q}|\omega_{n}|}
\label{10}
\end{align}
The main point is, that a corresponding $\chi_{QMC}\left(q,\tau\right)$
can be obtained directly from QMC, following preceding work employing the 
stochastic series expansion method \cite{grossjohann09}. This involves only the statistical
error, which is well controlled. Uncontrolled sources of error, due
to e.g. transformations to real or Matsubara frequencies, do not occur.
$\chi\left(q,\tau\right)$ is gauged against $\chi_{QMC}\left(q,\tau\right)$
by fitting $K_{q}$, $v_{q}$, and $\gamma_{q}$ at \emph{small} momentum,
while retaining $b$ and $c$ as given by bosonization. This is justified,
because the latter two constants do not enlarge the space of fitting-parameter,
as any modification of them can be absorbed into a renormalization
of $K_{q}$, $v_{q}$, and $\gamma_{q}$. Regarding the temperature
range, we confine ourselves to $T/J\leq0.25$. This is motivated by
the PE to $O(g^{3},T^{2})$ for thermodynamic properties to agree
rather well with QMC results up to $T/J\lesssim0.1$ \cite{johnston00b},
while for $T\gtrsim0.25$ the PE starts to fail significantly. 

Fig. \ref{fig:1} shows the result of the comparison of QMC with eqn. (\ref{10}) for the smallest
non-zero wave-vector $q=\pi/64$ of a 128-site system for two temperatures
$T/J=0.1$ and 0.25 allowing for three different choices of $\gamma_{q}$,
namely (i) $\gamma_{q,QMC}$ as optimized by fitting, (ii) $\gamma_{B}$
taken from the bosonization, and finally (iii) $\gamma_{q}=0$ forced
to be zero \cite{Mathematica}. The upper panel b) of this figure clearly
demonstrates, that QMC is inconsistent with $\gamma_{q}=0$ and that increasing
$\gamma_{q}$ above zero improves the quality of the fit. In particular
the best fit, i.e. for $\gamma_{q,QMC}$, is identical within
the standard deviation $2\sigma$ (error bar) to QMC for almost all
$\tau\in[0,\beta]$ at both temperatures. 
Yet, we find $\gamma_{q,QMC}>\gamma_{B}$, and moreover there are
{\em systematic} oscillatory deviations. While the latter seem a 
subdominant effect, which could be due to the on-shell approximation
in eqns. (\ref{5}) and (\ref{6}), these deviations may also indicate
relevant corrections to diffusion and should be
investigated in future studies. We emphasize the vertical scale
on panel b) of fig. \ref{fig:1} which demonstrates that high-precision
QMC is mandatory for the present analysis.
Fig. \ref{fig:1} is a central result of this
work. It shows that QMC is consistent with a dynamic structure factor
of the isotropic antiferromagnetic Heisenberg chain which is approximately diffusive at
intermediate temperatures in the long wave-length limit with a diffusion kernel $(1+c)v^2/(2\gamma_{q,QMC})$.
Any momentum dependence of $\gamma_{q,QMC}$, to be discussed later,
implies corrections to this diffusion. Next, and to further support
our approach, we will also discuss the Luttinger parameters we find.

\begin{table}[b]
\noindent \begin{centering}
\begin{tabular}{|c|c|c|c|c|c|}
\hline 
$T/J$  & $\chi_{q,QMC}/\chi_{0}$ & $\chi_{q,QMC}/\chi_{PE}$ &
$\gamma_{B}$\cite{Sirker09}  & $\gamma_{tTMRG}$\cite{Sirker09} 
& $\gamma_{q,QMC}$\tabularnewline
\hline 
0.1  & 1.0005 & 1.0032 & 0.0096  &  & 0.0191\tabularnewline
\hline 
0.25  & 1.0005 & 1.0248 & 0.0440  &  & 0.0511\tabularnewline
\hline 
0.2  &  &  & 0.0297  & 0.0190  & \tabularnewline
\hline
\end{tabular}
\par\end{centering}

\caption{\label{tab:1}Columns 2 and 3: Comparison of $\chi_{q,QMC}=K_{q}/
(2\pi v_{q}(1+c))$ from QMC at $q=\pi/64$ with $\chi_{0}$ from TBA
\cite{TheKlumperFiles}  and $\chi_{PE}$ from the l.h.s. of eqn. (\ref{8}).
Columns 4, 5, and 6 display $\gamma$ from bosonization, tTMRG, and QMC.}
\end{table}

\begin{figure}[t]
\begin{centering}
\includegraphics[scale=0.6]{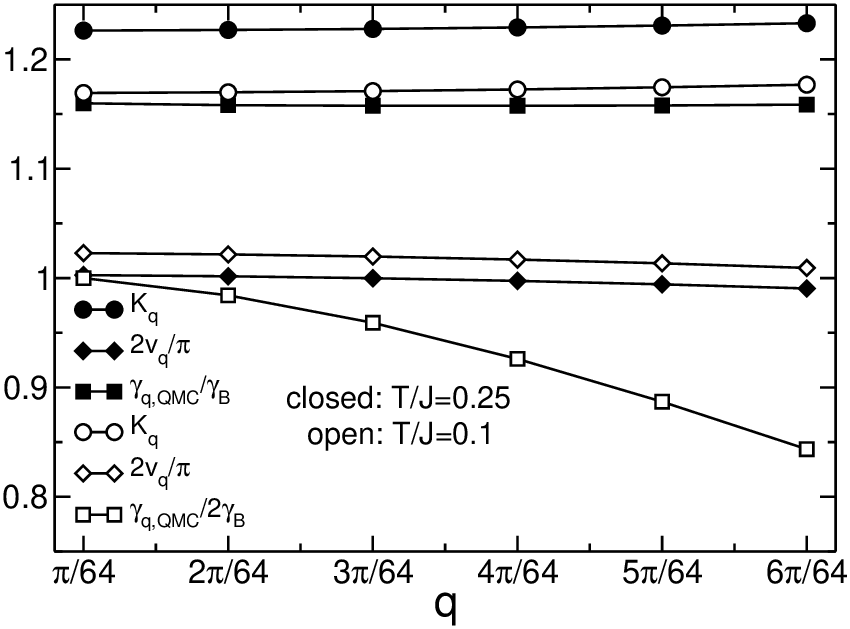}
\end{centering}

\caption{\label{fig:2}Momentum dependence of the renormalized Luttinger
parameter $K_q$, spinon velocity $v_q$, and scattering rate
$\gamma_{q,QMC}/\gamma_{B}$ for the first non-zero
six momenta on a 128 site system for two temperatures
$T/J=0.1$ (white symbols) and $0.25$ (black symbols). Note that
$\gamma_{q,QMC}/\gamma_{B}$ for $T=0.1$ has been
\emph{scaled by 2} to fit into the plot. }
\end{figure}

In table \ref{tab:1} we compare the parameters obtained from the
fit to QMC with results from TBA, PE and tTMRG. This table shows,
that $\chi_{q,QMC}=K_{q}/(2\pi v_{q}(1+c))$ at $q=\pi/64$ is in
excellent agreement with the isothermal susceptibility at $q=0$ from
the TBA for \emph{both} temperatures which we have studied. This result
should not be confused with the well know agreement between static
QMC and TBA for the isothermal susceptibility \cite{johnston00b},
but rather it is a satisfying consistency check for
our approach. In fact, fitting the imaginary-time transform of an
{\em approximate} $\chi(q,\omega)$, i.e. eqn. (\ref{5}), to QMC
could require values for $K_q$, $v_q$, and $\gamma_q$ which
deviate from exactly known values for these quantities
on a scale which is unrelated to the error $2\sigma$ of the QMC. 
As will be shown later the variation of $K_{q}$ and $v_{q}$
with momentum is very weak as $q\ll1$, i.e. we expect no relevant
change for $\chi_{q,QMC}$ as $q\rightarrow0$. Yet we
are tempted to point out, that $\chi_{q=\pi/64,QMC}$ in table
\ref{tab:1} is barely larger than $\chi_0$, which is consistent
with the momentum dependence for the exact $\chi_q$. The fact that
$\chi_{q,QMC}/\chi_{PE}>1$ and is increasing as $T$ increases, evidences
that $\chi_{PE}$ on the l.h.s. of eqn. (\ref{8}) increasingly underestimates
the TBA result as $T$ increases beyond $T/J\gtrsim0.1$. 
In fig. \ref{fig:1} we have shown, that $\gamma_{q,QMC}\neq
\gamma_{B}$. Yet, table \ref{tab:1} demonstrates that $\gamma_{q,QMC}$
and $\gamma_{B}$ are comparable to within factors of order 2. Most
important, the relaxation rate $\gamma_{q,QMC}$ we find is much
larger than the width of the two-spinon continuum, yet, very small
compared to temperature $\gamma_{q,QMC}\ll T$. We note, that fits
to tTMRG \cite{Sirker09} at $T/J=0.2$, lead to $\gamma_{tTMRG}/
\gamma_{B} \approx0.64$.

Next we discuss the momentum dependence. Fig. \ref{fig:2} displays
all three fit parameters $K_{q}$, $v_{q}$ and $\gamma_{q,QMC}$
versus the first six non-zero momenta and the two temperatures $T/J=0.1$
and 0.25 which have also been considered in fig. \ref{fig:2}. $v_{q}$
and $\gamma_{q,QMC}$ have been normalized to their values
given by bosonization, i.e. $\pi/2$ and $\gamma_{B}$. Obviously
all momentum variations are very smooth and rather weak. 
As can be seen from this figure,
most of the renormalization of the ratio $K_{q}/v_{q}$ from its bare
value of $2/\pi$ stems from $K_{q}>1$. 
The spinon velocity $v_{q}$ deviates slightly from $\pi/2$, however only
to within $O(1\%)$. As discussed in the previous paragraph, this is necessary
to obtain an optimum fit of the QMC to the approximation eqn. (\ref{10}) and
does {\em not} imply that QMC is at variance with the bare spinon velocity.
$K_{q}$ displays a very weak upward curvature, while $v_{q}$
shows a small downward curvature. The latter can be understood in terms
of the $O(q^{2})$ corrections to the linear on-shell dispersion $\omega(q)$
which are not contained in bosonization. The combined momentum dependence
of $K_{q}/v_{q}$ leads to the expected increase of the static susceptibility
with $q$. Finally, $\gamma_{q,QMC}/\gamma_{B}$ also displays a
weak momentum dependence which is larger for $T/J=0.1$. The latter may
signal the onset of finite size effects. In fact, $\gamma_{q,QMC}\neq 0$
implies a length scale $l$ of order $O(v/(2\gamma_{q,QMC}))$ for the
regular current relaxation. $l$ is less than the system size for both
temperatures studied. Yet, $128/l\approx 9$ for $T/J=0.25$ and
and $128/l\approx 3$ for $T/J=0.1$.
With momentum dependence, $\gamma$ as extracted from a
real-space quantity \cite{Sirker09} will differ from that obtained by
QMC at fixed small momenta.

\begin{figure}[t]
\noindent \begin{centering}
\includegraphics[scale=0.55]{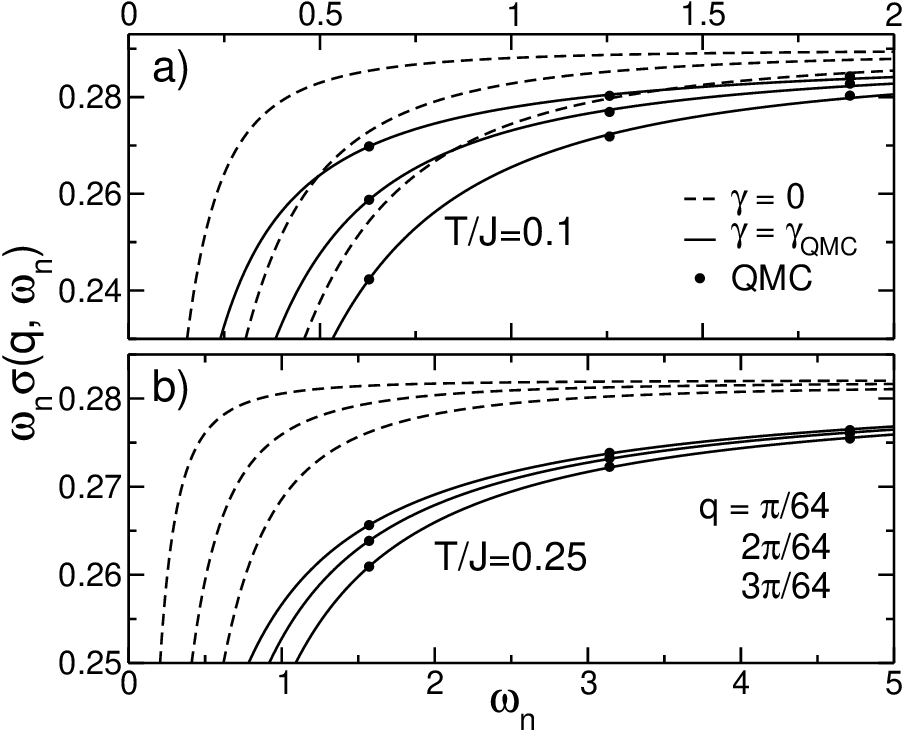}
\par\end{centering}

\caption{\label{fig:3}$\omega_{n}\sigma_{QMC}(q,\omega_{n})$ from QMC
for the first three non-zero Matsubara frequencies $\omega_{n}=2\pi nT$
and wave vectors $q=n\pi/64$, with $n=1,2,$ and 3 as compared to
$\omega_{n}^{2}\,\chi(q,\omega_{n})/q^{2}$ using eqn. (\ref{10})
with $\gamma=0$ (dashed) and $\gamma=\gamma_{q,QMC}$ (solid) on a 128 site
system for a) $T/J=0.1$ and b) 0.25. (See text
regarding statistical error.)}
\end{figure}

While the preceding has been exact up to the statistical error of
the QMC, we would like to conclude this work by speculating on the
line-shape of the regular part of the conductivity on the imaginary
frequency axis at $\omega_{n}=2\pi nT$. In principle this requires
a careful analysis of the error introduced by the Fourier transform
$\chi_{QMC}(q,\omega_{n})=\int_{0}^{1/T}\exp(i\omega_{n}
\tau)\chi_{QMC}(q,\tau)d\tau$. This error will increase 
as $\omega_{n}$ increases. Here we refrain from analyzing this, 
since our goal is merely to demonstrate to which extend
our QMC data discriminates between a conductivity with $\gamma=0$
and one with $\gamma=\gamma_{q,QMC}\neq0$. To this end fig.
\ref{fig:3} displays $\omega_{n}\,\sigma_{QMC}(q,\omega_{n})=
\omega_{n}^{2}\,\chi_{QMC}(q,\omega_{n})/q^{2}$
as compared to $\omega_{n}\,\sigma(q,\omega_{n})=
\omega_{n}^{2}\,\chi(q,\omega_{n})/q^{2}$
with $\chi(q,\omega_{n})$ taken from eqn. (\ref{10}) and with
$\gamma=0$ or $\gamma=\gamma_{q,QMC}$. Without any
further ado, this figure clearly demonstrates that
$\gamma=0$ in $\sigma(q,\omega_{n})$ from eqn. (\ref{7}) and 
(\ref{10}) is inconsistent with our QMC which however
agrees very well with $\sigma(q,\omega_{n})$ for
$\gamma=\gamma_{q,QMC}$ \cite{AGqmc}. This implies that
QMC is consistent with a Drude type of behavior
of the frequency dependence of the regular conductivity with
a relaxation rate $2\gamma_{q,QMC}$. While future
studies, may focus on finite size scaling, to perform the
limit of $q\rightarrow 0$, as required in eqn. (\ref{7}), this is
beyond the scope of the present analysis.

In conclusion QMC is consistent with spin dynamics
of the isotropic 1D Heisenberg antiferromagnet which is primarily
diffusive in the long wave-length limit and at intermediate temperatures,
implying a {\em regular} part of the spin conductivity with
a finite relaxation rate $\gamma\ll T$. This corroborates recent
findings by bosonization and tTMRG. Our analysis does not allow
conclusions on the pending open questions on the Drude weight
at $\Delta=1$, yet based on the numerical evidence for $D(T>0)>0$,
our findings may open up the intriguing possibility of
a finite temperature dynamical spin conductivity of the isotropic
Heisenberg model which comprises of both, a finite Drude
weight and a regular part with a very large mean free path at low
temperatures. Future analysis should focus on the relevance of
corrections beyond the on-shell approximation, on the case
$\Delta<1$, and on higher temperatures $T\gtrsim J$.

We are indebted to F. Heidrich-Meisner and R. G. Pereira for valuable
comments. Work supported in part by the Deutsche Forschungsgemeinschaft
through Grant No. BR 1084/6-1, FOR912 and by the National Science
Foundation under Grant No. PHY05-51164.


\begin{thebibliography}{25}
\bibitem{AGqmc}For a different QMC approach, not suited to clarify
the role of $\gamma$ in $\sigma(q,\omega_{n})$ see ref. \cite{Sirker09}
\end{thebibliography}
\end{document}